\documentclass[12pt]{iopart}

\usepackage[latin1]{inputenc}   
\usepackage[francais]{babel}    
\usepackage{fancyhdr}       
\usepackage{geometry}       
\usepackage{color}
\usepackage{graphicx}       

\newcommand{\be}{\begin{equation}}
\newcommand{\ee}{\end{equation}}
\newcommand{\bc}{\begin{center}}
\newcommand{\ec}{\end{center}}
\newcommand{\ei}{\end{itemize}}
\newcommand{\und}[1]{\underline{#1}}
\newcommand{\cls}{\mathcal{B}\mathcal{C}_{\mathcal S}}
\newcommand{\uus}{\underline{\underline{\sigma}}}

\begin{document}


\title  {Diffusion in pores and its dependence on boundary
conditions} \vspace{1cm}

\author{A. SAUGEY\dag, L. JOLY\ddag, C. YBERT\ddag, J.L.
BARRAT\ddag, L. BOCQUET\ddag }

\address{\dag  Laboratoire de Tribologie et Dynamique des
Syst\`emes, Ecole Centrale de Lyon and CNRS,
        36 Avenue Guy de Collongues, BP163, 69134 Ecully Cedex,
        France}
\address{\ddag  Laboratoire de Physique de la Mati\`ere Condens\'ee
et Nanostructures,
         Universit\'e Claude Bernard Lyon I and CNRS, 6 rue Amp\`ere, 69622 Villeurbanne Cedex, France}

\begin{abstract} We study the influence of the boundary
conditions at the solid liquid interface on diffusion in a
confined fluid. Using an hydrodynamic approach, we compute
numerical estimates for the diffusion of a particle confined
between two planes. Partial slip  is shown to significantly
influence the diffusion coefficient near a wall. Analytical
expressions are derived in the low and high confinement limits,
and are in good agreement with numerical results. These
calculations indicate that diffusion of tagged particles could be
used as a sensitive probe of the solid-liquid boundary conditions.
\end{abstract}

\section{Introduction}
The no-slip boundary condition for a fluid near a solid surface is
still under debate \cite{granick,cottin05}. At the macroscopic
scale, the no slip boundary condition is a consequence of the
microscopic roughness \cite{richardson}. On the nanometer scale
however partial slip is possible, and has indeed  been measured experimentally
\cite{lauga_review}.
This issue, which is important both fundamentally
and for the conception of  microfluidic devices, has motivated a
number of theoretical \cite{bocquet94,andrienko} and numerical
studies \cite{cieplak} . These studies  have highlighted the
influence of the fluid-wall interaction and pressure on the slippage
\cite{bocquet99,sokhan,watanabe}. While chemical
heterogeneities and surface roughness are expected to decrease
slippage \cite{pit}, surfaces with special geometries can exhibit
a "super-hydrophobic" state with a strongly increased slippage at the surface
\cite{cottin03,cottin04} that makes fluid dynamics at solids
surfaces very sensitive to surface imperfections. Such effects
have been evidenced using micro-engineered surfaces in reference
\cite{rothstein}. 

Slippage is usually accounted for in terms of an extrapolation
length, the so-called slip length, here denoted as $\delta$
\cite{bocquet94}.  This is defined as the distance inside the
solid wall where the {\it extrapolated} flow profile vanishes.
More specifically this partial slip boundary condition is written,
for the tangential component, $v_t$, of the velocity as
\begin{equation}
v_t = \delta {\partial v_t \over {\partial n}}
\end{equation}
with $n$ the coordinate in the direction normal to the solid
surface. The precise value of this slip length and its dependence
on the physical and chemical characteristics of the surface have
been investigated in a number of recent experimental studies. In
particular, very different values for slip lengths -from a few
nanometers to microns - have been reported using  different
techniques (see e.g. \cite{lauga_review} for a review). Many of
these techniques are indirect (pressure drop measurements
\cite{watanabe,rothstein}, particle image velocimetry
\cite{tretheway}, fluorescence recovery \cite{pit}), or very
delicate (surface force apparatus \cite{cottin05} and Atomic force
microscopy \cite{craig}). Hence the development of complementary,
robust and non-intrusive techniques to investigate the dynamical
properties of the solid-liquid interface would provide valuable
counterparts of the previous results.

In this manuscript, we
discuss how the diffusion of tagged particles between walls is
affected by confinement, and how such measurements could be used
as a signature of the nature of the boundary conditions \cite{bech,almeras,joly}.

We develop a theoretical and numerical approach to
estimate the roles of confinement and slip on diffusion
constrained in a planar or cylindrical pore. We make use of a
classical hydrodynamic description, which is expected to be appropriate
for colloidal particles, and was previously shown to be also well
adapted for molecular diffusion \cite{bocquet94}. A numerical
approach is used for the general case, and analytical expressions
are derived in the high and low confinement regimes.

\section{Hydrodynamic estimate of the diffusion constant}
\label{sec:hydro}
The quantity of interest is the mean diffusion coefficient of colloidal tracers
averaged over the measurement volume. The latter is limited here by the presence of
the confining walls.
From a theoretical point of
view, a moving particle $\mathcal{P}$ is subjected to a friction
force proportional to its velocity. When the motion takes place in
a confined volume, the mobility $\mu$  depends on the boundary
conditions at the confining walls.  For a velocity $\und U$
parallel to the boundary, the diffusion coefficient $D_\parallel$
is  given by Einstein's relation \cite{Einstein1905} \be
D_\parallel=\mu k_B T \label{eqn:einstein} \ee For a particle
moving between two flat walls separated by a distance $H$,
 $D_\parallel$ is a function of the
particle radius $a$, the height $H$ and the position $z$ of the
particle respective to the walls (in the following $z$ will be
measured by taking the origin at the midplane). The average
diffusion coefficient in the direction parallel to the walls is
\begin{eqnarray}
    \langle D_\parallel \rangle=\frac{1}{H-2a}\int_a^{H-a} D_\parallel(z)dz
    \label{Dave}
\end{eqnarray}

The next step is to use the so called Stokes Einstein approach,
i.e. to estimate the friction force from hydrodynamics. At low
Reynolds number, the flow around the particle is governed by the
Stokes equations :
\begin{eqnarray}
    \eta\Delta \und{V}&=&\und{\nabla} P  \label{eqn:navierstokes} \\
    \und\nabla.\und{V}&=&0               \label{eqn:incompress}
\end{eqnarray}
where $\und V$ is the velocity field, $P$ the pressure field and
$\eta$ the viscosity of the fluid. The boundary conditions are
\begin{itemize} \item fluid at rest at infinity in the unconfined
directions : \be \left.\und{V}\right|_{\infty}~=~\und{0}
\label{eqn:clinfty} \ee \item no slip on the particle surface  :
\be \left.\und{V}\right|_{\mathcal P}~=~\und{U} \label{eqn:clpart}
\ee \item partial slip on solid walls, expressed by parallel
$\left.\und{V}_\parallel\right|_{\mathcal S}$ and perpendicular
$\left.\und{V}_\perp\right|_{\mathcal S}$ velocities and slip
length $\delta$ :
\begin{eqnarray}
\delta\left.\nabla_\perp\und{V}_\parallel\right|_{\mathcal S}-\left.
\und{V}_\parallel\right|_{\mathcal S}&=&\und{0}  \label{eqn:clsolpara}\\
\left.\und{V}_\perp\right|_{\mathcal S} &=& 0
\label{eqn:clsolperp}
\end{eqnarray}
This condition is written $\cls(\und{V})=\und{0}$.
\ei

The friction force experienced by the particle is then: \be
\und{F}~=~\int\!\!\!\int_{\partial \mathcal P} \uus.\und{dS}
\label{eqn:forceexpr} \ee where
$\uus=-P~\und{\und{I}}+\eta\left(\und{\nabla}~\und{U}+^{t}\und{\nabla}~\und{U}\right)$
is the stress tensor in the fluid.

In the next sections we provide various solutions to this boundary problem :
we first start with a numerical ``exact'' solution of these equations; the latter will be used
subsequently as a reference solution for the approximate analytical solutions obtained
under various assumptions.

\section{Numerical Estimates}
\label{sec:num}

In this section we first start with a numerical solution of the previous equations,
Eqs. (\ref{eqn:navierstokes}) to (\ref{eqn:clsolperp}).
This set of equations was solved numerically with the  \emph{FEMLAB}$^\copyright$
software. A finite
domain of size $2L \times 2L \times H$  around the particle was
considered. The size $L$ was chosen large enough compared to $H$ to avoid
finite size problems~: typical values are $L/a=20$ for small H,
$L=3H$ otherwise. Space and time symmetries were taken into
account to reduce the meshed domain for faster computations.

From a technical point of view, the \emph{FEMLAB} fluid dynamics module
solves the bulk equations as
\begin{eqnarray}
    \und \nabla.\und{\und\sigma}=\und 0  \\
    \und \nabla.\und{V} = 0
\end{eqnarray}
with the stress tensor $\uus$ given above. Boundary conditions are imposed
according to Eqs. (\ref{eqn:clpart}) and (\ref{eqn:clsolpara})-(\ref{eqn:clsolperp}).

Once the flow field has been  obtained in this geometry, the force
is computed according to equation (\ref{eqn:forceexpr}). Note that
this way of computing the force requires a fine mesh since a
differentiation of the velocity field is performed. A better
approach would consist in using a weak constraint formulation so
that velocity and force are simultaneously computed on the
surface. However, such an approach is time and memory consuming
and was not used in this work to keep computational time within
reasonable bounds.

Typical results are shown in figure \ref{fig:figure2} where the profile of the {\it local}
parallel diffusion coefficient is plotted as a function of the altitude in the confining slab.
As a case study, we consider the situation in which one of the two walls has
a non zero slip length, while the no-slip boundary condition is
applied at the second wall. Near the no slip wall, diffusion
decreases from its bulk value as a result of the viscous friction
and high velocity gradient in the fluid between the particle and
the wall. This well known phenomenon \cite{happel} is easily
explained in term of an image particle (see next section). For a
no-slip wall, the image particle moves in the opposite direction
thus increasing the viscous force acting on the particle. Near a
partially slipping wall diffusion increases from the no slip case
and can even be higher than the bulk value. In the limit
$\delta\to\infty$, diffusion reaches a high value that can be
estimated using the image particle approach, with the image moving
in the same direction as the particle \cite{bech}.

We now turn to analytical approximate solutions of the Stokes
equation in the previous geometry.
\section{Analytical expressions in the low confinement (large gap) limit }
\label{sec:low} When the particle is small compared to
confinement height, an iterative reflection method can be developed, leading to an analytical
expression for the friction force.

In the present work we use this approach in the presence of a
single, slipping, wall. Then, summing over the forces due to each
wall yields an approximate result for the average diffusion
coefficient. A summary of the method is given here, and details
are discussed in Appendix \ref{app:solv1}.

\subsection{Reflection method with a single, slipping, wall}
The reflection
method is an iterative approach \cite{happel}, in which the velocity field $\und
V$ is expanded in the form
\begin{eqnarray}
    \und V & = & \und{V}^0+\und{V}^1+\und{V}^2+\und{V}^3 + \dots
\end{eqnarray}
with each $\und{V}^n$ field  satisfying the  bulk
 equations (\ref{eqn:navierstokes})-(\ref{eqn:incompress}). The zero order field,
$\und V_0$,  is chosen as the flow field around a sphere moving in the bulk :
\be
\und{V}^0(r,z)~=~\frac{3}{4}aU\left(\frac{2}{r}\und\nabla(z)-\und\nabla(\frac{z}{r})
                +\frac{a^2}{3}\und\nabla(\frac{z}{r^3})\right)
\label{eqn:chpbulk} \ee
with $a$ the radius of the sphere. This velocity field
satisfies the boundary equations on the
particle (equation (\ref{eqn:clpart})) and at infinity (equation
(\ref{eqn:clinfty})). The method consists in determining
$\und{V}^1$ field such that $\und{V}^0+\und{V}^1$ satisfies the
boundary conditions at infinity and on the solid walls
(\ref{eqn:clsolpara})-(\ref{eqn:clsolperp}) :
\[\left\{
\begin{array}{rcl}
\cls(\und{V}^1)               &=&-\cls(\und{V}^0)          \\
\left.\und{V}^1\right|_\infty &=&\und{0}
\end{array}\right.
\]
Now, at this level of approximation, the boundary condition on the
particle $\mathcal P$ is no longer satisfied by
$\und{V}^0+\und{V}^1$ and the next order $\und{V}^2$ is defined
from the reflection of $\und{V}^1$ on the particle as~:
\[
\left\{
\begin{array}{rcl}
\left.\und{V}^2\right|_{\mathcal P}  &=&-\left.\und{V}^1\right|_{\mathcal P} \\
\left.\und{V}^2\right|_{\infty}      &=&\und{0}
\end{array}\right.
\]
The higher moments of the velocity field, $\und{V}^n$, are  built by applying iteratively the
boundary condition on the particle and on the flat walls.

\subsection{Viscous force acting on the particle : a single wall }

The friction
force experienced by the particle is the sum of individual
contributions $\und{F}^n$ of each reflection : \be
\und{F}^n~=~\int\!\!\!\int_{\partial \mathcal P} \uus^n.\und{dS}
~=~\int\!\!\!\int\!\!\!\int_{\mathcal P} \und\nabla.\uus^n dV \ee
where $\uus^n$ is the stress tensor in the fluid. For odd
reflections, the velocity is regular in the volume of the particle.
The momentum equation gives $\und\nabla.\uus^n=0$ in the domain
occupied by $\mathcal P$ and the integral vanishes. For even
reflections, the Lorentz reciprocal theorem \cite{happel} gives
algebraically, {\it in the limit of small particles},
$F^{n+2}~=~\frac{V^{n+1}_{\mathcal O}}{V^{n-1}_{\mathcal O}}F^n$, where
 $V^{n}_{\mathcal O}$ is defined as the value of the velocity field
 at the center of the particle.
 One thus obtains
 \be \und F~=~-\und{F}^0 \sum_{k=0}^\infty
\left[-\frac{V_{\mathcal
O}^{1}}{U}\right]^k~=~-\frac{\und{F}^0}{1+\frac{V_{\mathcal
O}^{1}}{U}} \label{eqn:fricforce} \ee
with $\und{F}^0=6\pi\eta a
\und U$. As a consequence, only the velocity of the first
reflected field at the center of the particle $V_{\mathcal O}^{1}$
is needed to determine mobility and diffusion coefficient. The
calculation  of this field  is  described in appendix A.

Equations (\ref{eqn:fricforce}) and (\ref{eqn:vitcentre}) give the
force acting on a particle moving along a single planar wall as a
function of the radius of the particle $a$, the distance from the wall $l$
and the slip length $\delta$ :
\begin{eqnarray}
    \und{F}_{1wall} & = & \frac{6 \pi \eta a }{1-\frac{a}{z}C\left[\frac{l}{\delta}\right]}\und U
    \label{SW}
\end{eqnarray}
where the function $C$ is defined as
\be
C\left[{y}\right]=-\frac{3}{32}y^2-\frac{9}{32}y-\frac{3}{8}+\left(\frac{3}{32}y^3+\frac{3}{8}y^2+\frac{3}{8}y\right)E(y)+\frac{3}{2}yE(2y)
\ee
with $E(y)=e^yE_1(y)$ and $E_1(y)$ is the exponential integral function, defined as
$E_1(z)=\int_z^\infty dt ~e^{-t}/t $ \cite{Abram2}.

When $\delta \to 0$ (no slip condition), one recovers the well known value
\begin{eqnarray}
    C\left[\frac{l}{\delta}\right]\to \frac{9}{16}
\end{eqnarray}
derived from the method of the image particle \cite{happel} : as
mentioned above, diffusion decreases near a no slip wall. In the
limit $\delta \to \infty$ (full slip condition),
\begin{eqnarray}
    C\left[\frac{l}{\delta}\right]\to -\frac{3}{8}
\end{eqnarray}
and the presence of the wall reduces the friction force, i.e.
diffusion increases, as measured experimentally \cite{bech}.

Comparisons with numerical simulations using FEMLAB$^\copyright$ are shown in
figure \ref{fig:figure3}. Results are in good agreement down to
very small distances $l/a=1.5$. At large distance, one recovers
the bulk diffusion value as expected.

Moreover, simple and practical approximations can be obtained for the mobility in the limit where the
distance to the wall, $l$, is large compared to the slip length $\delta$.
Indeed an asymptotic expansion of $C\left[{y}\right]$ allows to obtain
\be
C\left[{y}\right]= {9\over 16} {1\over {1+{1\over y} + {\cal O}\left[ {1\over y^2} \right]}}
\ee
This gives the approximate following form for the friction coefficient
\begin{eqnarray}
    \und{F}_{1wall} & \simeq & \frac{6 \pi \eta a }{1-{9\over 16}\frac{a}{l+\delta}}\und U
    \label{SWa}
\end{eqnarray}
This approximation amounts to replace the distance to the wall $l$ by $l+\delta$,
where the physical meaning of the slip length in terms of an extrapolation length
appears quite clearly in this limit.
In practice, note that the expression in Eq. (\ref{SWa}) leads to values which are
within $5\%$ to the explicit result in Eq. (\ref{SW}) as soon as $l/\delta>0.5$ !

After completing this work, we became aware of a similar
calculation by Lauga and Squires \cite{laugasquires} who computed
the viscous force on a spherical particle close to a wall, using
the same reflection method. The use of the "small particle"
approximation corresponds to computing the flow in response to a
force applied to  a point-like particle, and can be shown to
involve errors of order $(a/h)^3$ \cite{laugasquires}.

\subsubsection{Local diffusivity in a confined geometry}

In order to compute the friction coefficient for a particle confined between two planar walls,
we make the further assumption that each wall contributes independently to the
 shift in the friction force from its bulk value~:
\begin{eqnarray}
    \und{F}_{2walls}= \und{F}_{1wall}(z,\delta)+\und{F}_{1wall}(H-z,\delta)- \und{F}_{bulk}
    \label{F2wall}
\end{eqnarray}
where $H$ is the distance between the two walls and here $z$ denotes the distance
to the bottom wall.
The Einstein equation then yields for the parallel diffusion coefficient at a height $l$~:
\begin{eqnarray}
    D_\parallel=\frac{k_B T}{6 \pi \eta a U}\frac{1}{\frac{1}{1-\frac{a}{z}C\left[\frac{z}{\delta}\right]}
    +\frac{1}{1-\frac{a}{H-z}C\left[\frac{H-z}{\delta}\right]}-1}
\label{eqn:diff2walls}
\end{eqnarray}

This expression for the friction coefficient is checked against the ``exact'' numerical results
obtained using the FEMLAB software in  Figures \ref{fig:figure4} and \ref{fig:figure5}.
Over the various slip lengths $\delta$ and confinement gap $H$, the agreement is found to
be quite good, within $6\%$ as long as the confinement is no too strong ($h/2a>4$).
It can be observed that Eq. (\ref{eqn:diff2walls}) slightly underestimates the diffusion.

Note that a different approximation could be made for the
contribution of the two walls, by assuming that the mobility
(rather than its inverse) is affected independently by the two
walls \cite{faxen}. This approximation, however, turns out to be
less accurate than the previous approximation, in Eq. (\ref{eqn:diff2walls}).

When the particle is confined to  a cylindrical pore, a similar
method can be used and  provides an estimate of the viscous force
acting on the particle in the low confinement limit (see appendix A).
However, as opposed to the planar case, only a
numerical estimation of the reflected velocity at the center of
the sphere can be reached. An interesting difference between the
planar and the cylindrical case, is that in the latter case, except close to the
wall where the behavior is similar to a particle moving near a
planar wall, the force acting on a tracer particle is never smaller
than its bulk value even in the large slip length limit
(i.e. the diffusion is reduced). This is due to the necessary
recirculation of the fluid around the particle. More precisely,
boundary conditions at infinity (no flow) imposes the overall flow
rate on a section of the cylinder at zero. In the section centered
on the particle, a negative fluid flow rate has to balance the
positive flow rate of the particle $\pi a^2 U$. Hence the  viscous
force increases from the bulk value even when $\delta\to \infty$.
In the planar geometry, recirculation takes place at infinity in
the unconfined directions and this phenomenon does not take
place.
\section{Lubrication theory in the strong confinement limit}
\label{sec:high} When the confinement approaches the particle size
($H \simeq 2a$), the main part of the viscous force is expected to arise from
the high velocity gradient in the thin fluid films between each
wall and the particle.  In these regions, the fluid flow is
quasi-parallel to the wall and lubrication theory \cite{batchelor}
is expected to provide a good description of the velocity field. An approximation of the
force acting on the particle can then be derived.

We assume here that the fluid is confined between a fixed sphere and a solid
wall moving at velocity  $\und U=(U,0,0)$ (see figure \ref{fig:figure2}).
One approximates furthermore the sphere by a paraboloid $h(r)\simeq h_0+\frac{r^2}{2a}$,
with $r$ the distance to the axis of symmetry of the paraboloid.

Under the lubrication assumptions \cite{batchelor},
the Stokes equation (\ref{eqn:navierstokes}) for the velocity field, $\und
W=\left(W_x,W_y,W_z\right)$, reduces to~:
\begin{eqnarray}
\eta \frac{\partial^2 W_\parallel(x,y,z)}{\partial z^2} & = & \und \nabla_\parallel P(x,y)
\end{eqnarray}
The boundary conditions are written as
\begin{eqnarray}
    \frac{\partial W_x}{\partial z} = \frac{W_x-U}{\delta}  \\
    \frac{\partial W_y}{\partial z} = \frac{W_y}{\delta}
\end{eqnarray}
on the wall and
\begin{eqnarray}
    \und W & = & \und 0
\end{eqnarray}
on the particle.
These equations are easily integrated and using the conservation equation
$\und\nabla_\parallel.\und Q_\parallel=0$ for the flow rate,
 defined as $\und Q_\parallel=\int_0^{h(x,y)}\und W_\parallel dz$, one gets the following equation
for the pressure, $P$~:
\begin{eqnarray}
    -\frac{1}{12\eta}\und\nabla_\parallel\left(\frac{h^3(h+4\delta)}{h+\delta}\und\nabla_\parallel P \right)
    +\und U.\und\nabla_\parallel\left(\frac{h^2}{2(h+\delta)}\right) = 0
    \label{eqP}
\end{eqnarray}
A general solution for the pressure can be written in the form $P=P_\infty+\Pi(r)\cos(\theta)$, with
\{$r$,$\theta$\} the angular coordinates on the planar wall.
We could however not find an analytical solution for the previous differential equation. However
a ``heuristic solution'' could be found after some manipulation of the
differential equation in the form $\Pi(r)=\eta U ~r b[h(r)]$,
with 
\begin{eqnarray}
    b[h]=-\frac{6}{5h\delta}-\frac{9}{10}\frac{\ln(h)}{\delta^2}
         +\frac{4}{5}\frac{\ln(h+\delta)}{\delta^2}+\frac{1}{10}\frac{\ln(h+4\delta)}{\delta^2}
    \label{bb}
\end{eqnarray}
We refer to appendix \ref{app:lubri} for details of the
calculations leading to this result. The validity of this
approximate  expression for the pressure was checked by computing
numerically the solution of the full differential equation
(\ref{eqP}) using a simple ODE solver (Mathematica$^\copyright$).
One finds that the previous solution for the pressure differs from
the ``exact'' numerical one, from less than a few percents for
$\delta \in[0,R]$, $h_0\in[{R\over 20},R]$, and over the full
range of distance $r\in[0,\infty[$ (see figure \ref{fig:figure11}
in the appendix). Moreover, in the vanishing slip length limit,
$\delta \rightarrow 0$, the previous solution for $b[h]$ in Eq.
(\ref{bb}) reduces to the corresponding exact solution of the
differential equation, which can be easily obtained as
$b_0(h)=-\frac{6}{5h^2}$.

Using this previous heuristic solution as a good approximation for the pressure,
one may then write
the force balance along the x-direction applied on the volume of
fluid inside the cylinder $r<R_c$ (see figure \ref{fig:figure7})
as
\begin{eqnarray}
    F_{\mathcal P}=-\left(F_{wall}+\int_{r=R_c}P\und n.\und x dS \right)
    \label{FF}
\end{eqnarray}
At large scales $R_c\to \infty$, one may verify that the slip effect disappears~:
$P(\delta,r=R_C)\to P(\delta=0,r=R_C)$ and $\int_{r=R_c}PdS$ is
independent of $\delta$ and the dependence of the friction force acting on the particle
$P$ come from the $R_C\rightarrow \infty$ limit of $F_wall$.

A second difficulty however arises with the lubrication
calculation : whatever the slip length $\delta$, the friction
force on the wall, $F_{wall}=\int_{wall}\eta \nabla_\perp W_x dS$,
is found to be logarithmically divergent when $R_c\to \infty$
\cite{happel}. This can be easily verified by inserting in the
previous friction force expression the expression $W_x$ deduced
from the pressure field with Eq. (\ref{bb}) (see also Eq.
(\ref{Fdiv}) in appendix B). On the other hand, the {\it
difference} of friction forces, $\Delta
F_{wall}=F_{wall}[\delta]-F_{wall}[\delta=0]$, between the finite
slip length case and the no-slip case is found to take a finite
value, given in eq. (\ref{DF}). Note that $\Delta F_{\mathcal
P}=\Delta F_{wall}\equiv\Delta F$ (since the second term in Eq.
(\ref{FF}) is independent of $\delta$ in the $R_C\rightarrow
\infty$ limit).

One may however argue that this difference is a physically relevant quantity
since the slip effects mainly affect the flow in the region with strongest confinement.
One may indeed verify that at the lubrication level, the flows with partial slip
reduces to the flow with the no-slip boundary condition in the region far from
closest contact.

We have plotted in Figure \ref{fig:figure8} (left) the result for
$\Delta F=F_{wall}[\delta]-F_{wall}[\delta=0]$ (normalized by the bulk
value of the force on the particle $F_{\infty}=6\pi\eta a U$) as a function of
the minimum gap $h_0$ between the sphere and the wall. This result
is compared to the FEMLAB calculations in the same geometry. As expected
an agreement is found in the small gap limit, where the
lubrication approximation is expected to be valid.

Pursuing this calculation, a diffusion coefficient can be obtained. First
the friction coefficient on the particle situated at a distance $l$ from the
wall (with slip length $\delta$) can be estimated at this level of approximation
by adding to the previous $\Delta F$ the value of the friction force
$F_{wall}[l,\delta=0]$ computed in the same configuration for a no-slip wall.
Then the friction coefficient for the particle confined between two partially
slipping wall is estimated by adding the effects of the walls on the friction
coefficient, according to Eq. (\ref{F2wall}). The mobility is finally evaluated by
the inverse of the friction coefficient. This procedure is applied in Figure \ref{fig:figure8}
(right), where the numerical (FEMLAB) result has been used for the
no-slip friction force $F_{wall}[l,\delta=0]$. Here the diffusion coefficient is computed
only for the situation where the sphere is at the center of the slab ($z=H/2$).
This result is compared to ``exact'' results obtained using a full FEMLAB calculation for
the particle confined between two partially slipping walls.
Again, the lubrication approximation only yields a correct agreement in the small gap region,
and works better for small slip lengths. When the slip length increases,
lubrication theory overestimates the mobility : In this case,
important contributions to the viscous force are coming  from
areas far from the confined zones, which are not properly described
within the lubrication approach.

The lubrication approach has therefore a quite limited range of
application (in the very confined region) but the solution
obtained is complementary to the low confinement results which
works in the large gap limits.

\section{Averaged diffusivity and conclusions}

We are now in a position to compute the averaged diffusion coefficient over
the confined slab, $\langle D_\parallel \rangle$ defined
in Eq. (\ref{Dave}). We consider a geometry where one wall is characterized
by a no-slip boundary condition, while a partial slip boundary condition,
with a slip length $\delta$ applies on the other. This configuration is chosen
as to mimic the experimental geometry \cite{joly}.

Results are shown in figures  \ref{fig:figure9} and \ref{fig:figure10} for
various values of the  confinement $H$ and of the slip length
$\delta$.

 Analytical results obtained in the low confinement
approximation, combined with the assumption of independent wall
contribution, reproduce quite well the trends of the numerical
computations.  Figure \ref{fig:figure9} shows that the analytical
estimate slightly underestimates diffusion in the low confinement
limit, and tends to overestimate it at strong confinements.

In
order to observe a significant dependence of diffusion on the slip
length, two conditions are required. The particle size should not
be much larger than the slip length, and a sufficiently strong confinement
is required. With typical values of $H\simeq 4a$, variations
of the average diffusion constant of typically 5\% to 10\% would
be expected if the slip length is changed between $0.1a$ and $a$.

These results therefore suggest that diffusion measurements are
quite sensitive to boundary conditions on the solid substrate. This
opens new routes to measure slip length on the basis of the
thermal motion of colloidal tracers \cite{joly}.


\vskip1cm
{\bf Acknowledgments}
L.B. thanks Yannick Almeras, with whom this work was initiated.
\setcounter{section}{0}
\setcounter{subsection}{0}
\setcounter{equation}{0}
\renewcommand{\theequation}{\Alph{section}.\textsf{\arabic{equation}}}
\renewcommand{\thesubsection}{\Alph{section}.\arabic{subsection}}

\addtocounter{section}{1}
\section*{Appendix~\Alph{section}: First reflected field $\und{V}^1$.}
\label{app:solv1}
\subsection{Planar geometry}
For a particle moving at a distance $l$ from a planar surface, the
general form of $\und{V}^1$ satisfying equations
(\ref{eqn:navierstokes})-(\ref{eqn:incompress}) and
(\ref{eqn:clinfty}) is given by \cite{morse} :
\begin{eqnarray}
V_x^1~=~\frac{1}{2\pi}\int_0^\infty\!\!\!\int_0^{2\pi}
        \left[   \left(2-\cos^2(u)(k|z|+1)\right)\Theta^1
        +ik\cos(u)\Upsilon^1 +zk^2\cos^2(u)\Xi^1 \right]\nonumber\\[-2mm]
\times e^{ik(x\cos(u)+y\sin(u))-k|z|}dkdu
\end{eqnarray}
\begin{eqnarray}
V_y^1~=~\frac{1}{2\pi}\int_0^\infty\!\!\!\int_0^{2\pi}
i\sin(u)\left[   i\cos(u)(k|z|+1)\Theta^1
+k\Upsilon^1         -izk^2\cos(u)\Xi^1  \right]\nonumber\\[-2mm]
\times e^{ik(x\cos(u)+y\sin(u))-k|z|}dkdu
\end{eqnarray}
\begin{eqnarray}
V_z^1~=~\frac{1}{2\pi}\int_0^\infty\!\!\!\int_0^{2\pi}
        \left[  -izk\cos(u)\Theta^1
         -k\Upsilon^1         +ik\cos(u)(k|z|+1)\Xi^1\right]\nonumber\\[-2mm]
\times e^{ik(x\cos(u)+y\sin(u))-k|z|}dkdu
\end{eqnarray}
where $\Theta^1$, $\Upsilon^1$ et $\Xi^1$ are functions of
$k$ and $u$ and $(x,y,z)$ are cartesian coordinates centered
on the particle with $x$ along the particle velocity and $z$ normal to the wall.

The initial field $\und{V}^0$ is written in this form with
($\Theta^0(k,u)=\frac{3}{4}aU$, $\Upsilon^0(k,u)=
-\frac{1}{4}a^3Uik\cos(u)$, $\Xi^0(k,u)=0$). In the limit of a
small particle, $\Upsilon^0(k,u)<<\Theta^0(k,u)$.

Functions ($\Theta^1$, $\Upsilon^1$, $\Xi^1$) are determined
as the unique solution of $\cls(\und{V}^0+\und{V}^1)=\und{0}$ :
\begin{eqnarray}
\Theta^1    & = & \frac{3}{4}\frac { k\delta-1}{k\delta+1}aUe^{-2kl}  \\
\Upsilon^1  & = &  \frac{3i}{2}\frac{\left( 2lk^{2}\delta^2+k^2l^{2}\delta+k\delta l-\delta+kl^2-l\right)}{2k^2\delta^{2}+1+3k\delta} Ua\cos(u)e^{-2kl}  \\
\Xi^1       & = &  -\frac{3}{2}\frac{\left(k\delta l+\delta+l \right)}{2k^2\delta^2+1+3k\delta} aUe^{-2kl}
\end{eqnarray}
Integration of $V_x^1(0,0,0)$ gives
\begin{eqnarray}
V_{\mathcal O}^{1}=-\frac{a}{l}C \left[\frac{l}{\delta}\right]U
\label{eqn:vitcentre}
\end{eqnarray}
with  $C\left[{y}\right]=-\frac{3}{32}y^2-\frac{9}{32}y-\frac{3}{8}+\left(\frac{3}{32}y^3+\frac{3}{8}y^2+\frac{3}{8}y\right)E(y)+\frac{3}{2}yE(2y)$. $E(y)=e^yEi(1,y)$ and $Ei(1,y)$ is the exponential integral function.

\subsection{Cylindrical geometry}
For a particle moving in a cylinder, the general solution for the
reflected field $\und{V}^1$ in $(r,\phi,z)$ cylindrical
coordinates is
\begin{eqnarray}
\und V^1(r,\phi,z)&=&\frac{3aU}{2\pi}\sum_{k=-\infty}^{+\infty}\int_0^{+\infty}d\lambda
\left(\begin{array}{c}
a_k(\lambda,r)\cos(k\phi)\sin(\lambda z)\\
b_k(\lambda,r)\sin(k\phi)\sin(\lambda z)\\
c_k(\lambda,r)\cos(k\phi)\cos(\lambda z)
\end{array}\right)_{(\und e_r,\und e_\phi,\und e_z)}\label{eqn:cylsolgen}\\
a_k(\lambda,r)&=&\frac{k}{\lambda r}\Omega^1_k(\lambda) I_k(\lambda r)+\Psi^1_k(\lambda) I_k'(\lambda r)+\lambda r\Pi^1_k(\lambda) I_k''(\lambda r)  \nonumber\\
b_k(\lambda,r)&=&-\Omega^1_k(\lambda)I_k'(\lambda r)-\frac{k}{\lambda r}\Psi^1_k(\lambda)I_k(\lambda r)-k\Pi^1_k(\lambda)I_k'(\lambda r)+\frac{k}{\lambda r}\Pi^1_k(\lambda)I_k(\lambda r) \nonumber\\
c_k(\lambda,r)&=&\Psi^1_k(\lambda)I_k(\lambda r)+\lambda r\Pi^1_k(\lambda)I_k'(\lambda r)+\Pi^1_k(\lambda)I_k(\lambda r) \nonumber\\
\end{eqnarray}
where $I_k$, $I_k'$ et $I_k''$ are the first order modified Bessel functions and their derivatives.

Bulk field $\und{V}^0$ for a particle moving along the cylinder axis, at
a distance $b$ from it, is expressed in such a form as :
\begin{eqnarray}
\und V^0(r,\phi,z)&=&\frac{3aU}{2\pi}\sum_{k=-\infty}^{+\infty}\int_0^{+\infty}d\lambda
\left(\begin{array}{c}
\alpha_k(\lambda,r)\cos(k\phi)\sin(\lambda z)\\
\beta_k(\lambda,r)\sin(k\phi)\sin(\lambda z)\\
\gamma_k(\lambda,r)\cos(k\phi)\cos(\lambda z)
\end{array}\right)_{(\und e_r,\und e_\phi,\und e_z)}\\
\alpha_k(\lambda,r)&=&\left(\lambda r+\frac{k^2}{\lambda r}\right)K_k(\lambda r)I_k(\lambda b)+\lambda bK_k'(\lambda r)I_k'(\lambda b)
\nonumber\\
\beta_k(\lambda,r) &=&-k\left( K_k'(\lambda r)I_k(\lambda b)+\frac{b}{r}  K_k(\lambda r)I_k'(\lambda b)                \right)
\nonumber\\
\gamma_k(\lambda,r)&=&2K_k(\lambda r)I_k(\lambda b)+\lambda r K_k'(\lambda r)I_k(\lambda b)+\lambda b K_k(\lambda r)I_k'(\lambda b) \nonumber\\
\end{eqnarray}
where $K_k$, $K_k'$ et $K_k''$ are the second order modified Bessel functions and their derivatives.

$\Omega^1_k(\lambda)$, $\Psi^1_k(\lambda)$ and $\Pi^1_k(\lambda)$
are the unique solution of \be \cls(\und{V}^0+\und{V}^1)=\und{0}
\ee on the cylindrical wall $r=R$. The analytical expression is
then used to compute numerically $V^1_{\mathcal P}$ as : \be
V^1_{\mathcal
O}=\frac{3aU}{2\pi}\!\!\!\sum_{k=-\infty}^{+\infty}\!\!\!\int_0^{+\infty}\!\!\!\left[
\left(\Psi_k(\lambda)+\Pi_k(\lambda)\right)I_k(\lambda b)+\mu b
\Pi_k(\lambda)I_k'(\lambda b)\right]d\lambda \ee

\addtocounter{section}{1}
\section*{Appendix~\Alph{section}: Lubrication approximation}
\label{app:lubri}
Using reduced variables
$r=\sqrt{2h_0a}\tilde{r}$, $h=h_0\tilde{h}$,
 $\delta=h_0\tilde{\delta}$ and $P=\eta U \frac{\sqrt{2h_0a}}{h_0^2}\tilde{p}$,
 mass conservation is (for compactness w remove the $\tilde{ }$ for the reduced variables):
\begin{eqnarray}
    -\und\nabla_\parallel\left(\frac{h^3(h+4\delta)}{12(h+\delta)}\und\nabla_\parallel p \right)
    +\und e_x.\und\nabla_\parallel\left(\frac{h^2}{2(h+\delta)}\right) = 0
    \label{eqn:redmass}
\end{eqnarray}
with $h(r)=1+r^2$.

Assuming $p(r)=p_\infty+rb(r)\cos(\theta)$, equation (\ref{eqn:redmass}) becomes
\begin{eqnarray}
    -\frac{\partial}{\partial r}\left[r\alpha(h(r))\frac{\partial}{\partial r}\left(rb(r)\right)\right]
     +\alpha(h(r))b(r)+2r\beta(h(r))=0
\end{eqnarray}
with $\alpha(h)=\frac{h^3(h+4\delta)}{12(h+\delta)}$ and $\beta(h)=\frac{h^2+2\delta h}{2(h+\delta)^2}$.

We could not find an exact solution for this equation. However in order to proceed further,
we have tried to construct in a heuristic way a good approximation to the solution
to avoid purely numerical solutions. We have proceeded as follows.
First $b$ is assumed to depend functionaly on $h(r)$, as $b[h(r)]$.
Expressing $p=P_{\infty}+x b[h]$ ($x=r \cos\theta$) in Eq. (\ref{eqn:redmass}), this
equation rewrites
\begin{eqnarray}
    -\und\nabla_\parallel\cdot\left(\alpha(h) \left(b(h) \und{e}_x + x {\partial b \over \partial h}
    \und\nabla_\parallel h\right) \right)
    +2 \beta(h) x= 0
    \label{eqn:redmass1}
\end{eqnarray}
with $\und\nabla_\parallel h=\{2x,2y\}$ in cartesian coordinates. An heuristic solution
is found by assuming $\alpha(h), \beta(h)$ and ${\partial b \over \partial h}$ as
constant the previous equation, which amounts to replace the previous equation by
\begin{eqnarray}
    -6 x {\partial b \over \partial h}   +2{\mathcal A} \beta(h) x= 0
    \label{eqn:redmass3}
\end{eqnarray}
The constant ${\mathcal A}$ is adjusted so that the exact no-slip solution of
Eq. (\ref{eqn:redmass}), $b_0(h)=-\frac{6}{5h^2}$, is recovered. The solution of the
Eq. (\ref{eqn:redmass3}) with ${\mathcal A}=6/5$ is
\begin{eqnarray}
    b(h)=-\frac{6}{5h\delta}-\frac{9}{10}\frac{\ln(h)}{\delta^2}
         +\frac{4}{5}\frac{\ln(h+\delta)}{\delta^2}+\frac{1}{10}\frac{\ln(h+4\delta)}{\delta^2}
         \label{BBapp}
\end{eqnarray}
which indeed reduces to the no-slip solution $b_0(h)=-\frac{6}{5h^2}$ when
$\delta\to0$. Note also that in the limit $h>>\delta$, one also recovers $b(h)\to b_0(h)$ :
the pressure is independent of $\delta$ far from the particle.

The reduced viscous force acting on the wall is
\begin{eqnarray}
    F_{wall}=\int \frac{\partial W_x}{\partial z}rdrd\theta
\end{eqnarray}
$W_x$ is determined from Stokes equation $\frac{\partial^2 W_x}{\partial z^2}=\frac{\partial p}{\partial x}$ along with the boundary conditions and yields
\begin{eqnarray}
    F_{wall}=3\pi\int_0^\infty \left[\frac{1}{6}(b(r)+\frac{\partial}{\partial r}(rb(r)))\frac{h^2}{h+\delta}
                                     +\frac{2}{3}\frac{1}{h+\delta}\right]rdr+Cste
     \label{Fdiv}
\end{eqnarray}
Deviation of the viscous force from $\delta=0$ case is, back with dimensionalized variables :
\begin{eqnarray}
    \Delta F = 6\pi\eta a U \int_0^\infty\left[\frac{h-1}{6}\left(b_0(h)-b(h)\frac{h(h+2\delta)}{(h+\delta)^2}\right) -\frac{2}{6}\frac{\delta}{h(h+\delta)}\right]dh
\end{eqnarray}
This expression can be exactly computed for the approximated $b(h)$ given above :
\begin{eqnarray}
    \Delta F &=& \frac{6\pi\eta a U }{360\delta^2}
    \left(36\delta-12\delta^2+10\pi^2\delta^2
    +54\delta   \ln\left(\frac{1}{\delta}\right)
    -54\delta^2 \ln\left(\frac{1}{\delta}\right) \right. \nonumber \\&&
    +27\delta^2 \ln\left(\frac{1}{\delta}\right)^2
    +3\delta^2\ln\left(\frac{1}{3\delta}\right)^2
    -24\ln\left(1+\delta\right)
    +134\delta \ln\left(1+\delta\right)   \nonumber \\&&
    -190\delta^2\ln\left(1+\delta\right)
    -24\delta^2 \ln\left(1+\delta\right)^2
    -54\delta   \ln\left(\frac{1+\delta}{\delta}\right)
    +54\delta^2 \ln\left(\frac{1+\delta}{\delta}\right)  \nonumber \\&&
    -3 \ln\left(1+4\delta\right)
    -26\delta   \ln\left(1+4\delta\right)
  -56\delta^2 \ln\left(1+4\delta\right) \nonumber \\&&
  -6\delta^2 \ln\left(1+\delta\right)\ln\left(1+4\delta\right)
    +6\delta^2  \ln\left(1+\delta\right)\ln\left(\frac{1+4\delta}{3\delta}\right) \nonumber \\&&
    \left.+54\delta^2 {\rm Li}_2\left[\frac{-1}{\delta}\right]
    +6\delta^2 {\rm Li}_2\left[\frac{-1-\delta}{3\delta}\right]
    \right)
    \label{DF}
\end{eqnarray}
with ${\rm Li}_2(x)$ the dilogarithm function defined as ${\rm Li}_2(z)=\sum_{k=1,\infty} z^k/k^2$
\cite{Abram}.

\bibliographystyle{unsrt}

\begin{thebibliography}{10}
\bibitem{Abram} Note that this definition slighlty changes from that in ref. \cite{Abram2} where
it is defined as $\tilde{Li}_2(z)=\sum_{k=1,\infty} (-1)^k(z-1)^k/k^2$.
\end{thebibliography}

\newpage
\bc {\bf \large FIGURE CAPTIONS}\ec\vspace{2cm}

{\bf Figure 1 :} Geometry of the present calculations. A tracer particle with radius $a$
diffuses in a slab with thickness $H$.

{\bf Figure 2 :} Numerical estimates of the reduced diffusion coefficient of a particle moving between a partially slipping wall ($\delta=1$ : full line, $\delta=100$ : dotted line) at $z=0$ and a no-slip wall at $z=H$, as a function of  the position of the particle. From left to right, $H/a= 3, 5, 8, 12, 17 and 22$.

{\bf Figure 3 :} Diffusion coefficient near a single planar wall as a function of the distance $l$, for various slip length. Numerical results (solid lines) are compared with the analytical solution (dashed line) in the low confinement limit $l>>a$. The slip length $\delta$ increases from bottom to top.

{\bf Figure 4 :} Local diffusivity computed using the approximate analytical
results Eq. (\ref{eqn:diff2walls}) for $\delta/a=10^{-1}$ (dahsed line), compared to the numerical
results (solid line). See figure \ref{fig:figure2} for details and notations.

{\bf Figure 5 :} Same as in Fig. \ref{fig:figure4} but for $\delta/a=10^{1}$. See figure \ref{fig:figure2} for details and notations.

{\bf Figure 6 :} Flow description in the lubrication limit in the thin confined film

{\bf Figure 7 :} Sketch of the force balance in the volume $r<R_c$.

{\bf Figure 8 :} Numerical test of the lubrication calculations : (left) plot of the friction force difference
$\Delta F=F_{wall}(\delta)-F_{wall}(\delta=0)$ normalized by the bulk value $F_\infty=6\pi \eta a U$,
for a single wall, as a function of the distance $l$ to the wall. The solid line is the FEMLAB
calculation, while the dashed line is the lubrication estimate; (right) Diffusion coefficient for a particle in the middle plane of the confined geometry between 2 identical partially slipping walls, in the high confinement limit $H/2a\sim1$. A good agreement between the numerical and lubrication
calculations is found when $\delta \to 0 $ and $H/2a \to 1$.

{\bf Figure 9 :} Mean diffusion coefficient between a no-slip wall and a partially slipping wall ($\delta$) as a function of the gap $H$ for various slip lengths $\delta$. Full line : numerical results, dashed line : low confinement approximation (average of Eq. (\ref{eqn:diff2walls})).

{\bf Figure 10 :} Same as figure \ref{fig:figure9}, but plotted as a function of slip length $\delta$
for fixed confinement $H$. Full line : numerical results, dashed line : low confinement approximation
(average of Eq. (\ref{eqn:diff2walls})).

{\bf Figure 11 :} Negative pressure $-\Pi(r)=-r b(r)$ rescaled  by
$P_0=\eta U/a$, as a function of the radial distance $r$. The
minimum gap $h_0$ between the sphere and the solid surface is
$h_0=0.1 a$ and the slip length is $\delta=a$. The solid line is
the numerical solution of the equation for the pressure using a
ODE solver (Mathematica $^\copyright$). The dashed line is the
approximate solution, Eq. (\ref{BBapp}), see text.
\clearpage
\begin{figure}\centering
\includegraphics[width=10cm]{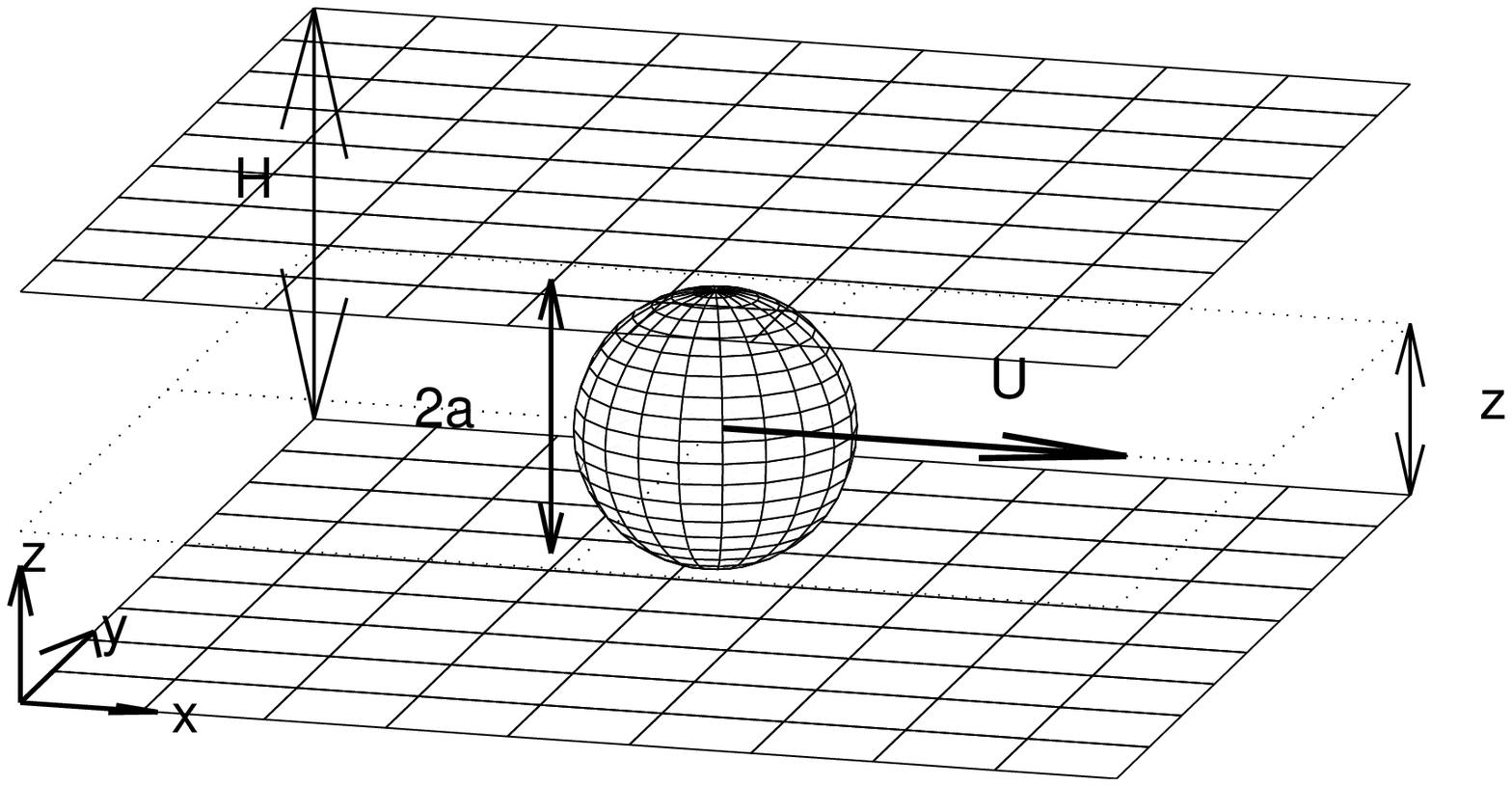}
\caption{}
\label{fig:figure1}
\end{figure}

\begin{figure}\centering
\includegraphics[width=10cm]{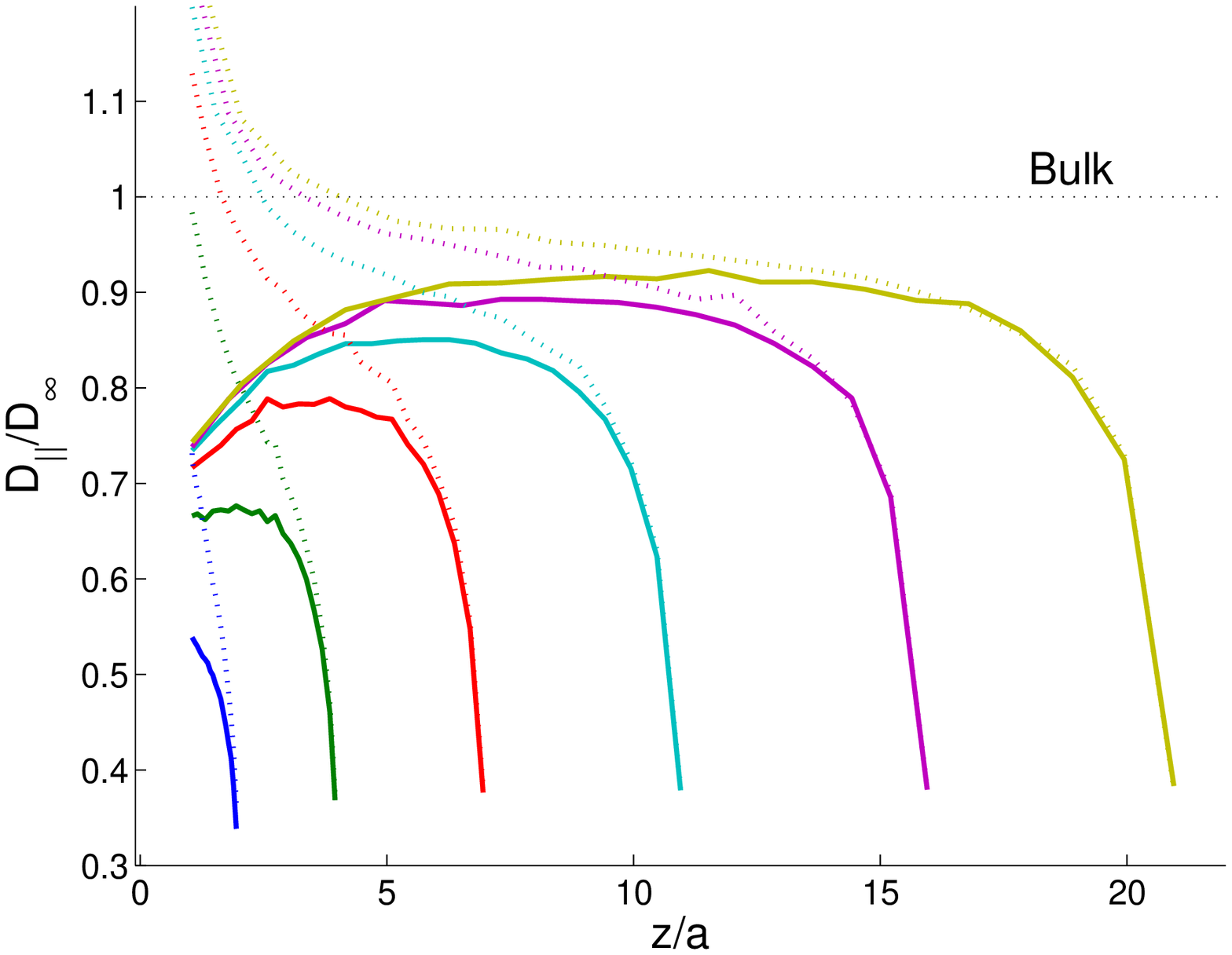}
\caption{}
\label{fig:figure2}
\end{figure}

\begin{figure}\centering
\includegraphics[width=10cm]{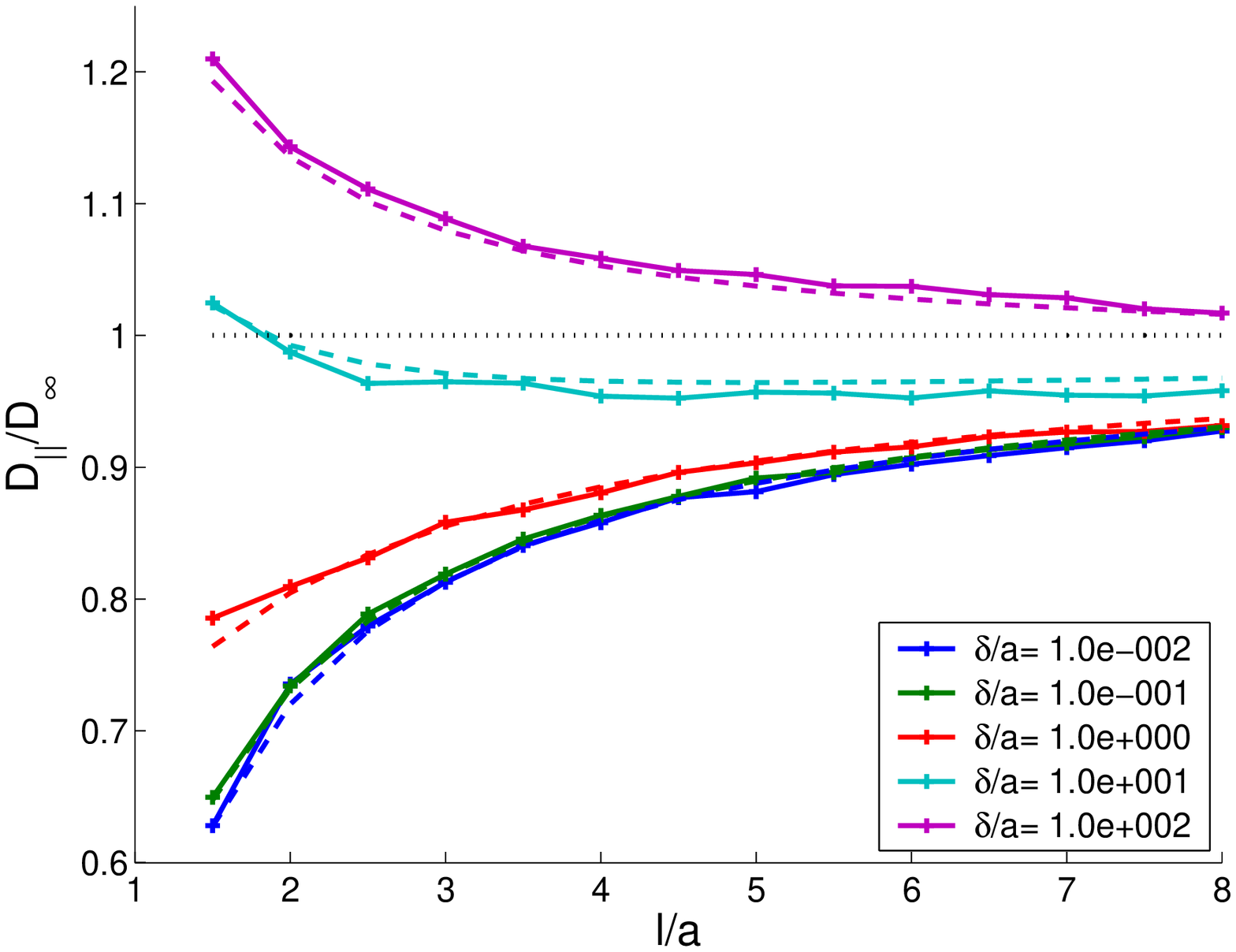}
\caption{}
\label{fig:figure3}
\end{figure}

\begin{figure}\centering
\includegraphics[width=10cm]{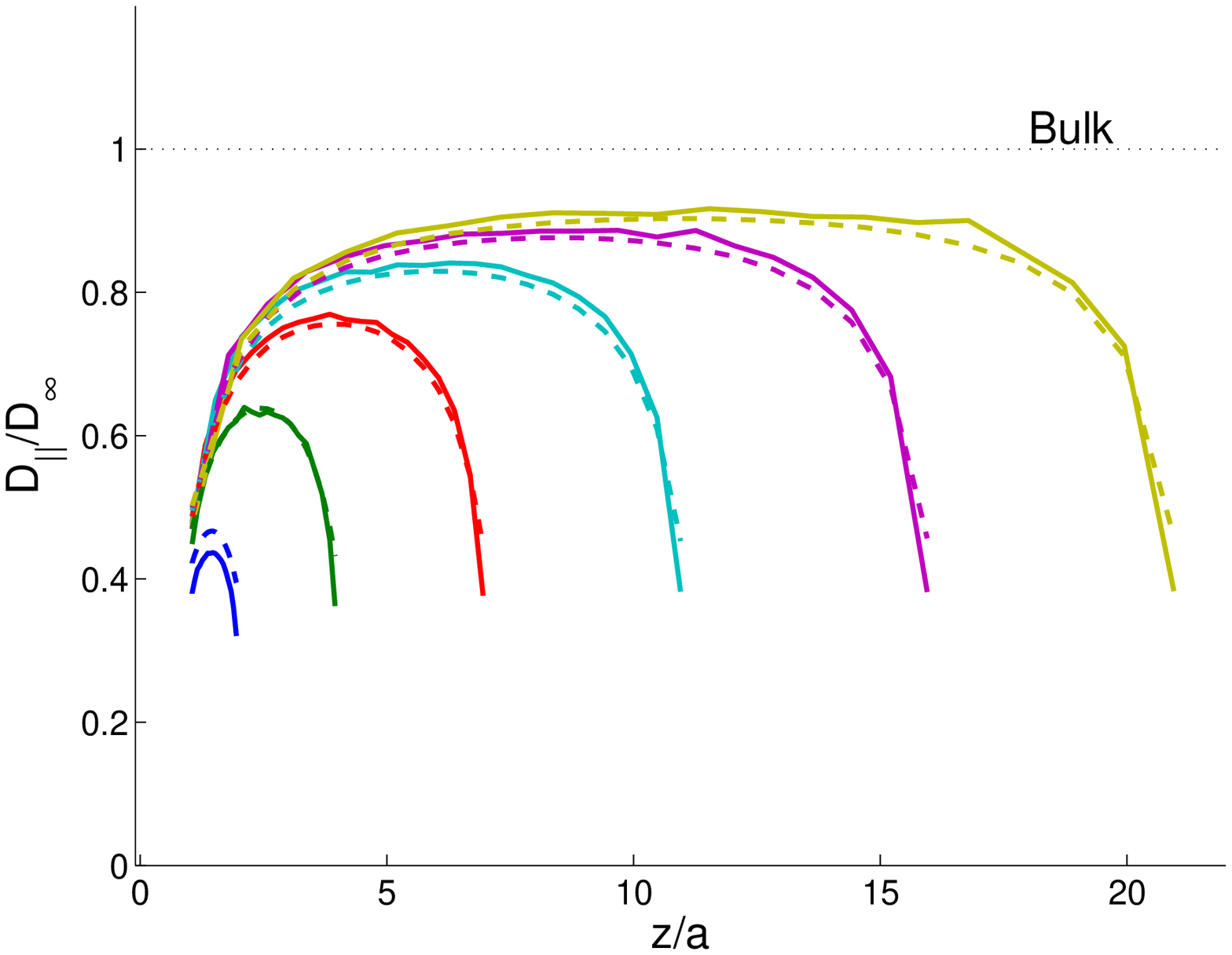}
\caption{}
\label{fig:figure4}
\end{figure}

\begin{figure}\centering
\includegraphics[width=10cm]{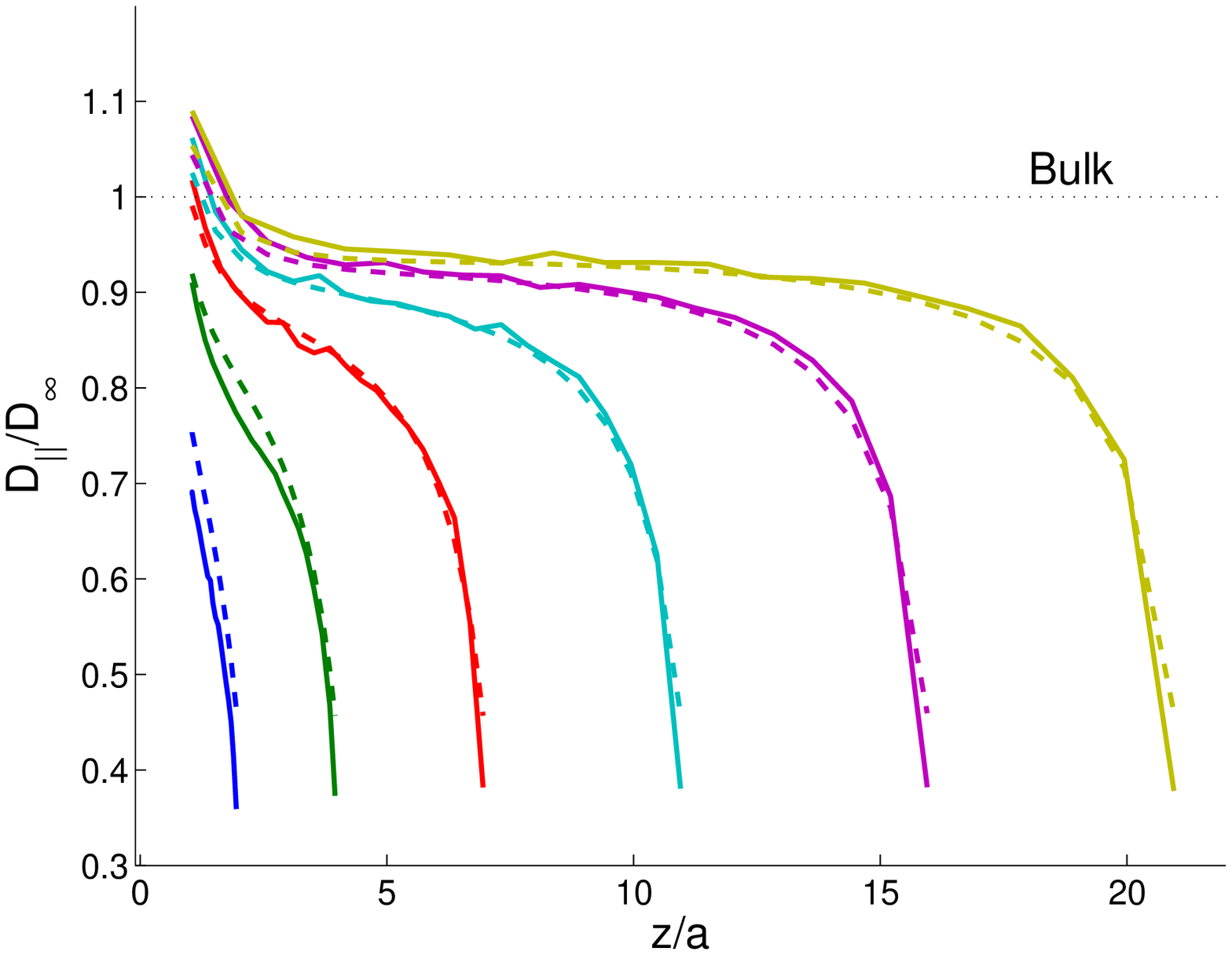}
\caption{}
\label{fig:figure5}
\end{figure}

\begin{figure}\centering
\includegraphics[width=10cm]{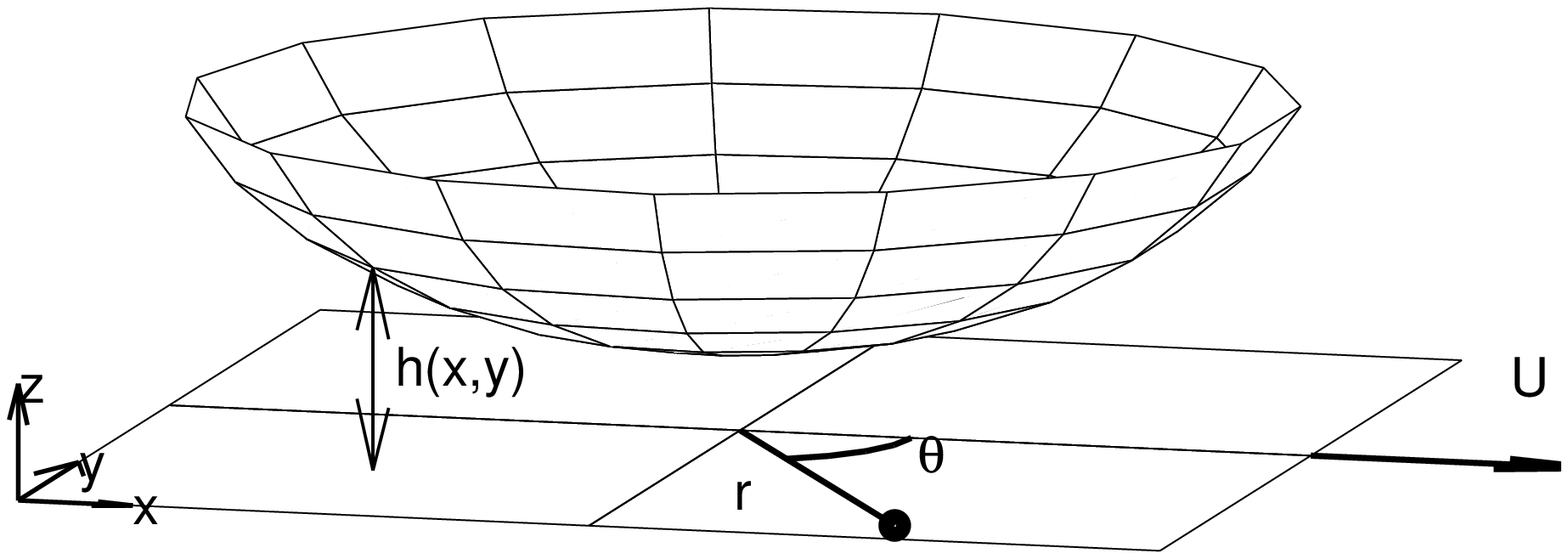}
\caption{}
\label{fig:figure6}
\end{figure}

\begin{figure}\centering
\includegraphics[width=10cm]{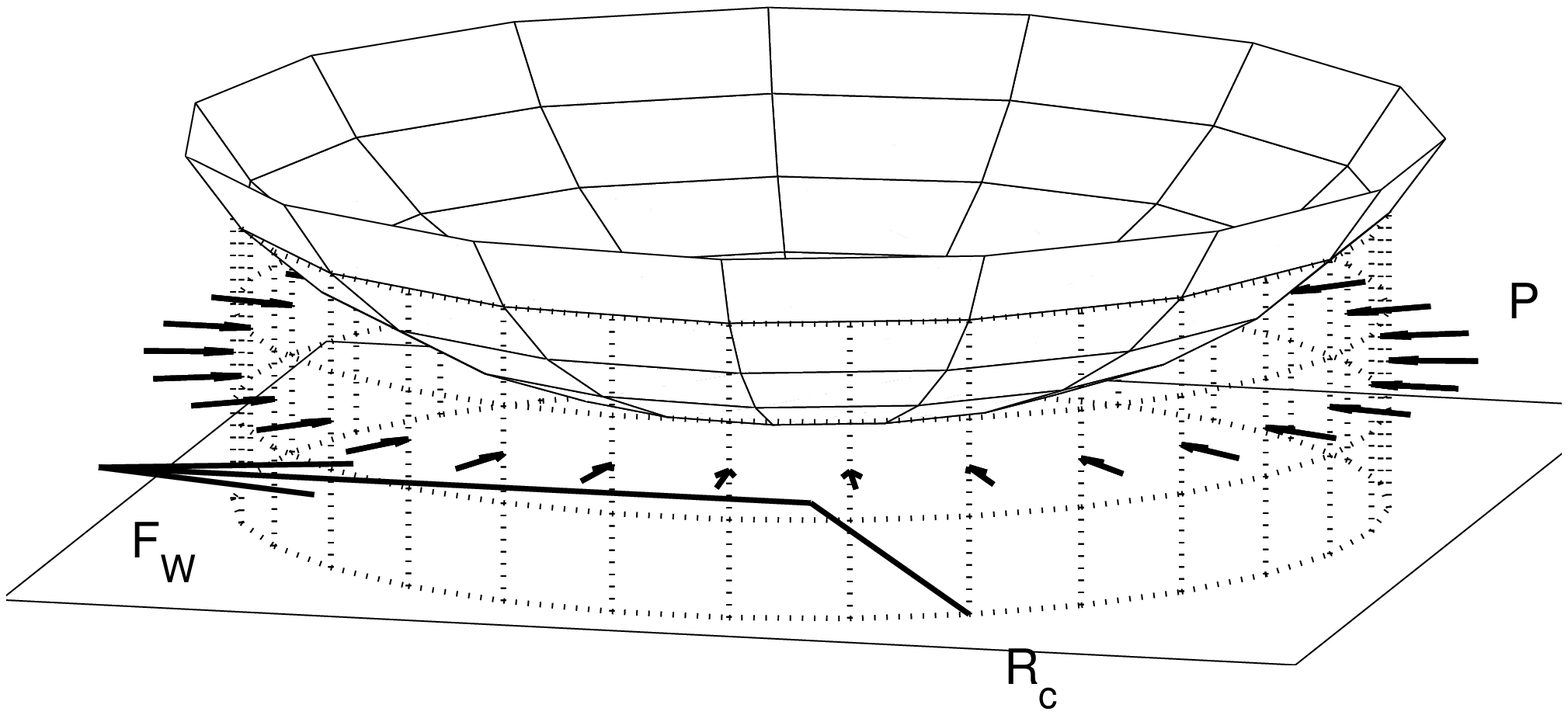}
\caption{}
\label{fig:figure7}
\end{figure}

\begin{figure}[h!]
\begin{center}
\includegraphics[width=7cm]{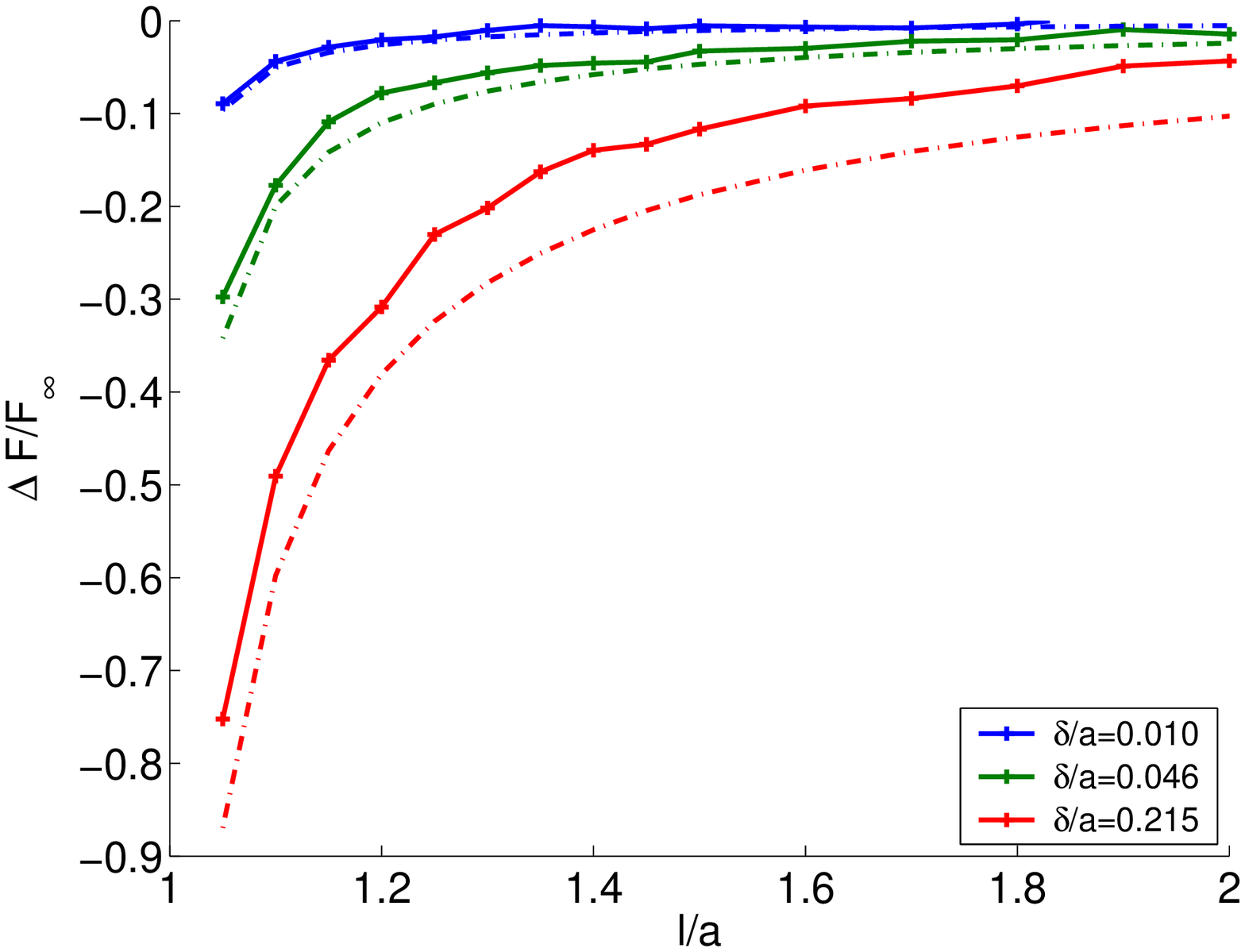}
\includegraphics[width=7cm]{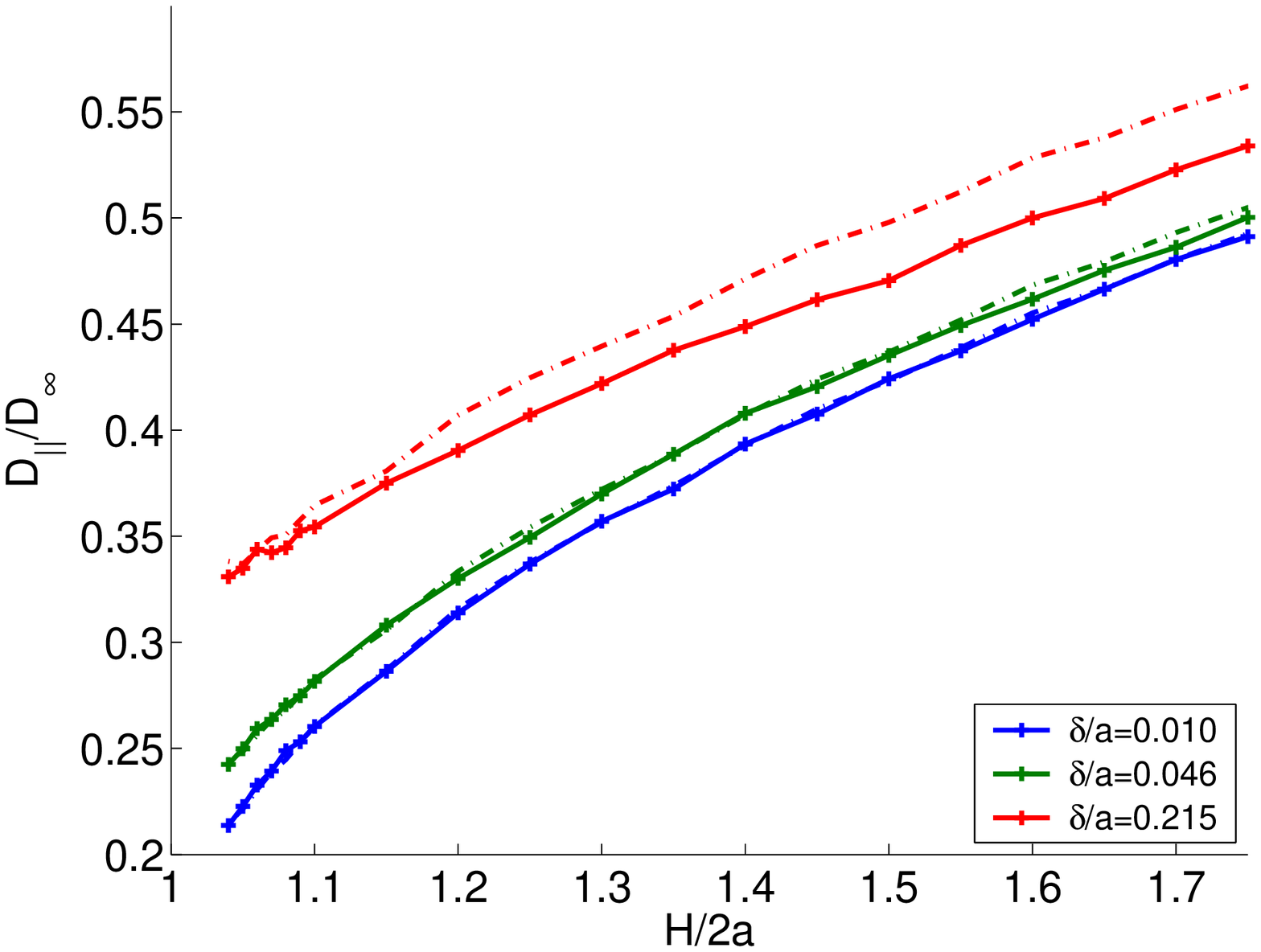}
\caption{}
\label{fig:figure8}
\end{center}
\end{figure}
\begin{figure}\centering
\includegraphics[width=10cm]{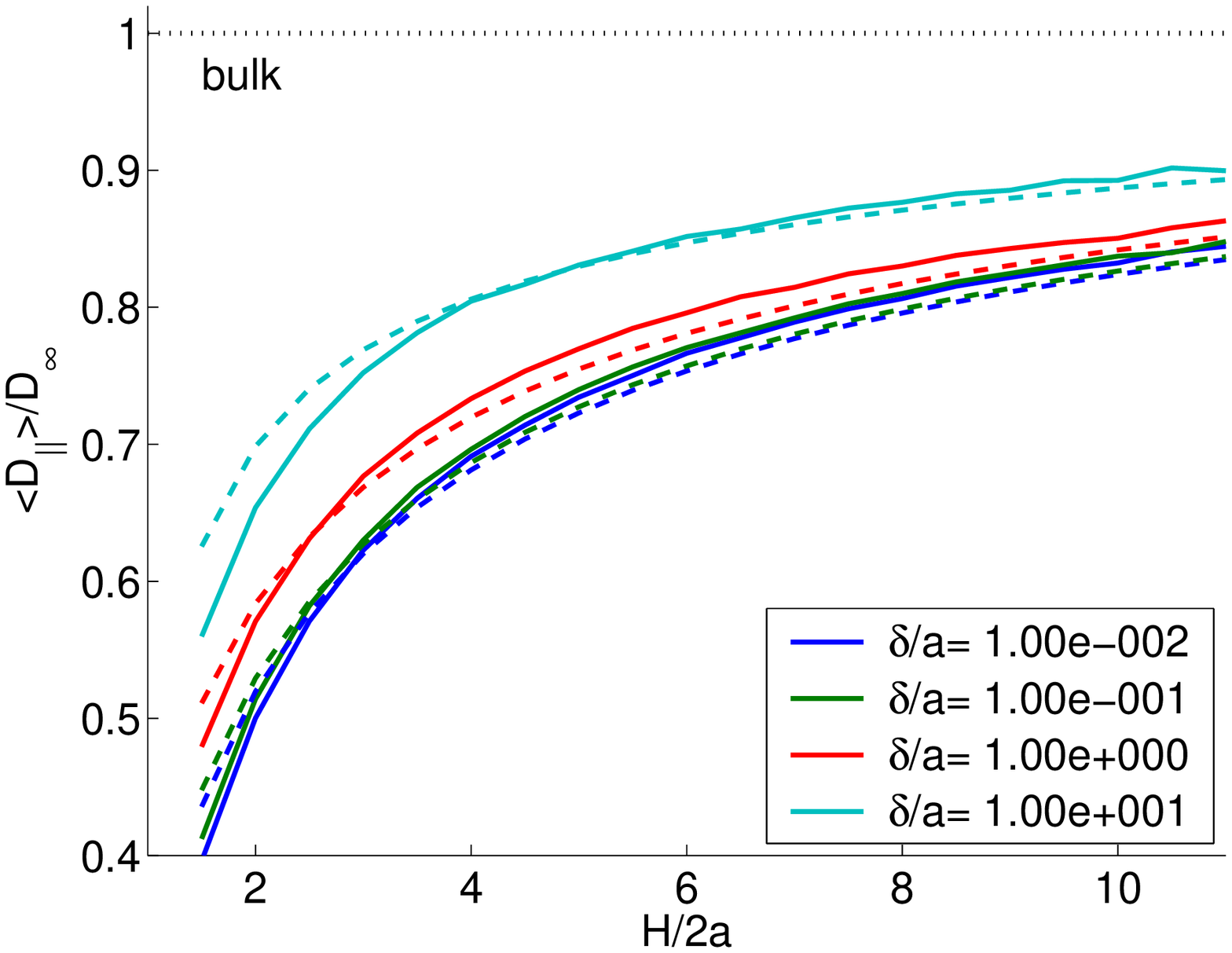}
\caption{}
\label{fig:figure9}
\end{figure}

\begin{figure}\centering
\includegraphics[width=10cm]{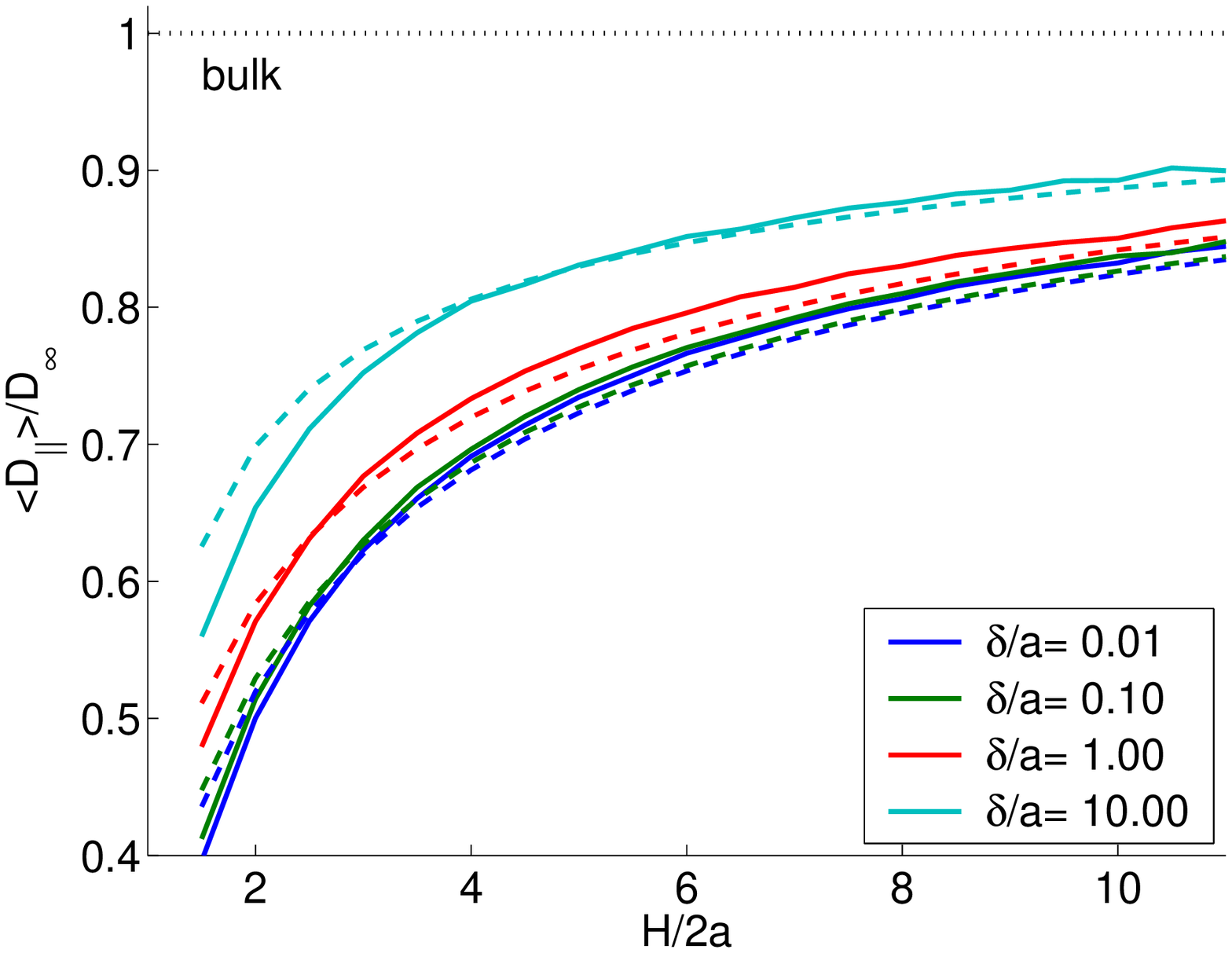}
\caption{}
\label{fig:figure10}
\end{figure}

\begin{figure}\centering
\includegraphics[width=10cm]{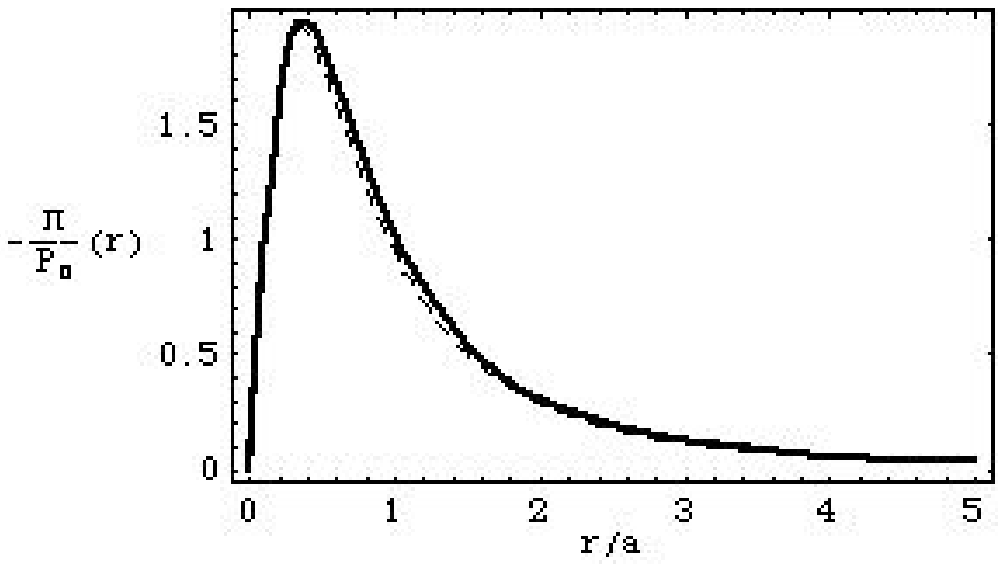}
\caption{}
\label{fig:figure11}
\end{figure}

\end{document}